\documentclass[aps,prc,superscriptaddress,showpacs,floatfix,notitlepage,twocolumn,nofootinbib,preprintnumbers]{revtex4-2}
    \usepackage{amsmath,graphicx,float,csquotes,mdframed,appendix}
\usepackage[colorlinks=true,linkcolor=blue,citecolor=blue, urlcolor=blue]{hyperref}

\newcommand{\nene}{$^{20}$Ne$^{20}$Ne}
\newcommand{\oooo}{$^{16}$O$^{16}$O}

\newcommand{\oo}{$^{16}$O}

\newcommand{\nee}{$^{20}$Ne}

\newcommand{\mpt}{$\langle p_T \rangle$}
\newcommand{\deltap}{$\Delta p_\text{max}$}
\newcommand{\deltar}{$\Delta r_\text{max}$}
\newcommand{\sigmazdc}{$\sigma_{NN}^\text{ZDC}$}

\begin{document}
 
\title{Transmutation of $^{16}$O and $^{20}$Ne at the Large Hadron Collider}

\author{Govert Nijs}
\affiliation{Theoretical Physics Department, CERN, CH-1211 Gen\`eve 23, Switzerland}

\author{Wilke van der Schee}
\affiliation{Theoretical Physics Department, CERN, CH-1211 Gen\`eve 23, Switzerland}
\affiliation{Institute for Theoretical Physics, Utrecht University, 3584 CC Utrecht, The Netherlands}
\affiliation{NIKHEF, Science Park 105, 1098 XG Amsterdam, The Netherlands}

\begin{abstract}

\noindent In July 2025 the Large Hadron Collider (LHC) collided $^{16}$O$^{16}$O and $^{20}$Ne$^{20}$Ne isotopes in a quest to understand the physics of ultrarelativistic light ion collisions. One particular feature is that there are many smaller isotopes with the exact same charge over mass ratio that potentially can be produced and contaminate the beam composition. Using the \emph{Trajectum} framework together with the GEMINI code we provide an estimate of the production cross-section and its consequences. A potential benefit could be the interesting measurement of the multiplicity and mean transverse momentum of $^{16}$O$^{4}$He and $^{20}$Ne$^{4}$He collisions.
\end{abstract}

\preprint{CERN-TH-2025-127}

\maketitle

\section{Introduction}

Ultrarelativistic light ion collisions are among the new exciting upcoming results coming from the LHC that among other things will enhance our understanding of relativistic hydrodynamcis at its limits and jet-medium interactions
in small droplets of quark-gluon plasma (QGP) \cite{Brewer:2021kiv,Grosse-Oetringhaus:2024bwr}\@. In this context there has been a recent effort to precisely understand the shapes of the isotopes \oo{} and \nee{}, which is both interesting and necessary to understand the upcoming collisions \cite{Giacalone:2024luz,Giacalone:2024ixe,Summerfield:2021oex,YuanyuanWang:2024sgp}.

In this short note we explore the possibility that the oxygen or neon beams transmutate into smaller isotopes that keep on circulating in the LHC. This is interesting by itself, a potential problem as a contamination of the original collisions, as well as an opportunity to learn about collisions with those contaminants.

\section{Model}

In this work, we use the \emph{Trajectum} framework, to which we have added a description of what happens to the spectators after the collision. For this we largely follow \cite{Liu:2022xlm,Liu:2023gun,Liu:2023qeq}, with a few modifications that will be highlighted.

\emph{Trajectum} uses a modified version of the T\raisebox{-0.5ex}{R}ENTo model for its initial conditions (see \cite{Nijs:2023yab})\@. In this model, certain nuclei are marked as \emph{wounded}, meaning that they participate in the collision. A naive definition of a spectator would be any nucleon that is not marked wounded. However, such a definition leads to an underestimation of the number of nucleons hitting the ZDC detectors in PbPb collisions \cite{ZDCtoappear}\@. The reason for this is that even if a nucleon participates in the interaction, it does not necessarily exchange a large amount of momentum. If the momentum exchange is small, the trajectory of a nucleon is essentially unchanged, meaning that for the purposes of hitting the ZDCs or not, these nucleons behave essentially the same as nucleons that are not marked as wounded.

To model this our nucleons can essentially be wounded (e.g.~participate in a hadronic collision) and be a spectator (e.g.~stay essentially around the rapidity of the beam)\@. We do this by introducing both the nucleon-nucleon cross-section $\sigma_{NN}$ (determining if a hadronic interaction occurs) and \sigmazdc{} (determining if a nucleon stays close to beam rapidity and is called a spectator)\@.

The spectators are then combined into clusters if they are within a space and momentum distance of \deltar{} and \deltap{} respectively. The sampling of the momenta of the nucleons follows \cite{Liu:2022xlm}, which takes the Fermi momentum based on the local nucleon density. In contrast to \cite{Liu:2022xlm}, where the nuclear structure model directly outputs a density that can be used to compute the Fermi momentum, we compute the density from the number of nucleons in a sphere immediately around each nucleon:
\begin{equation}
    p_\text{fermi}=(9\pi N_\text{nucl}/8)^{1/3}/r_\text{loc},
\end{equation}
where the radius $r_\text{loc}$ is a parameter in which to evaluate the local density and $N_\text{nucl}$ is the number of nucleons in a sphere of such radius.\footnote{If no other nucleons are present in the sphere we take $p_\text{fermi}=(9\pi/4)^{1/3}/s$ with $s$ the distance to the closest nucleon.}

The clusters are then fed into the GEMINI code, which decays excited isotopes with energy $E$ and angular momentum $J$ into  ground state isotopes until decays are forbidden or unlikely due to competition with photon emission \cite{Charity:1988zz,Charity:2010wk}\@. In \cite{Liu:2022xlm} there is a special treatment of deuteron, triton and $^3$He production, but for the purpose of this work we switched this off.

\begin{figure*}[t]
    \centering
    \includegraphics[width=18cm]{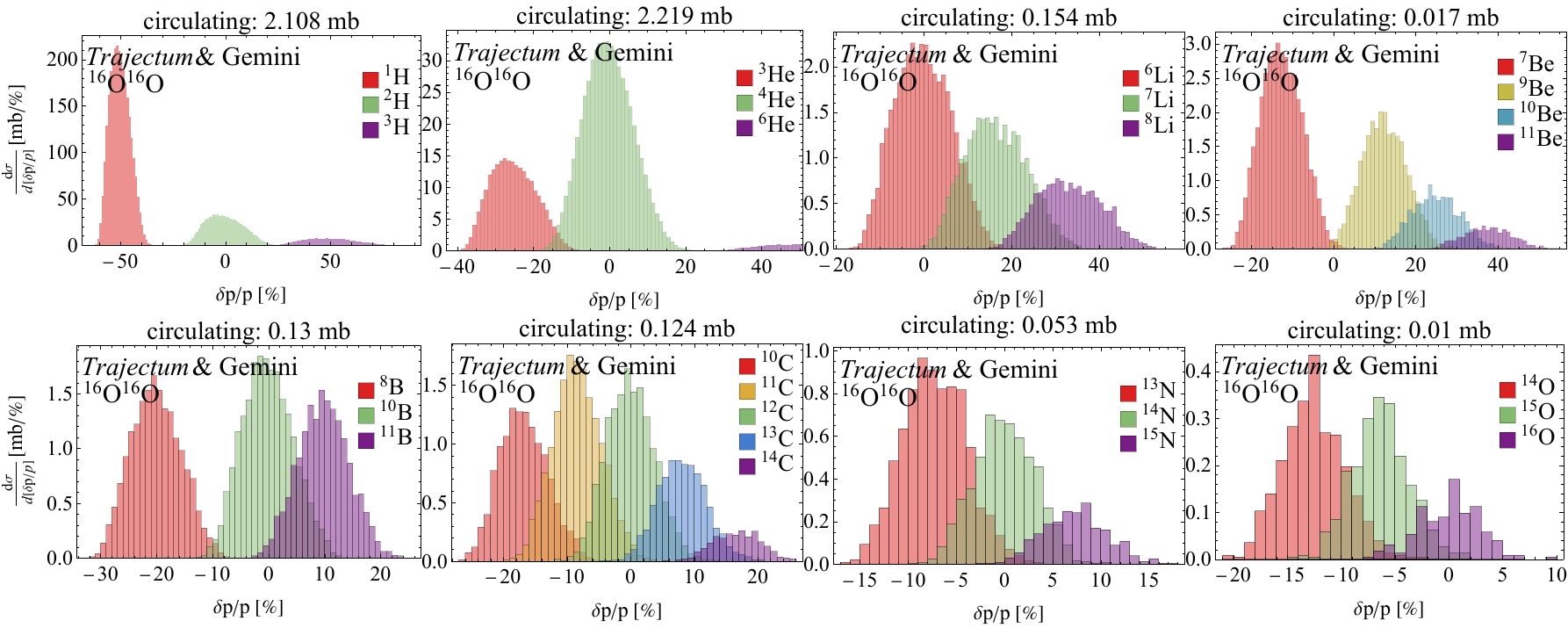}\\
    \includegraphics[width=18cm]{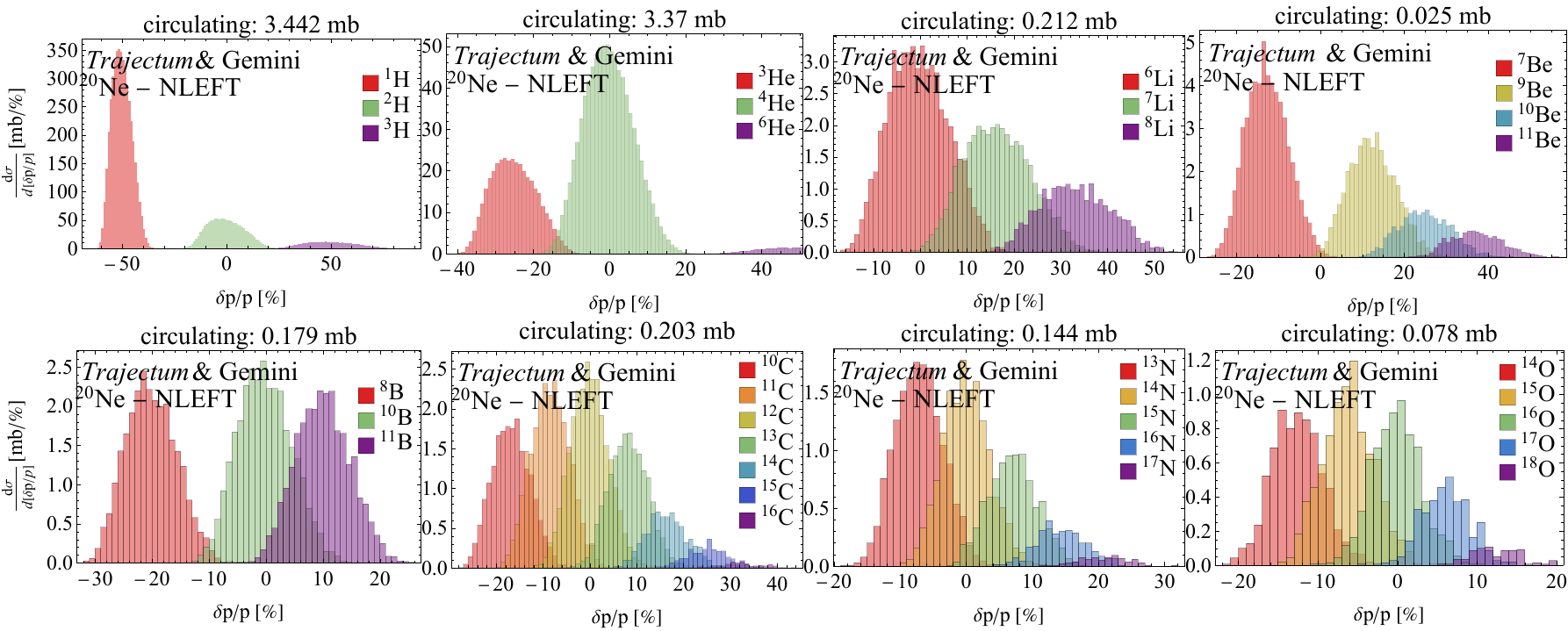}\\
    \caption{We show cross-sections of isotope productions as a function of momentum (relative to their `beam' momentum) by hadronic interactions for \oooo{} (top two rows) and \nene{} collisions (bottom two rows) for isotopes that have a lifetime greater than one second. We assume that isotopes with $|\delta p/p|<0.034\%$ can keep circulating in the LHC\@. The total hadronic \oooo{} and \nene{} cross-section is about $1.42\,$b and $1.85\,$b respectively for our model. The production of helium and other isotopes is enhanced by about 14\% more than a naive scaling with the cross-section would have given. }
    \label{fig:oxygentransmutation}
\end{figure*}


For this study we take $r_\text{loc}=2.51\,$fm, \deltar{}$=2.98\,$fm, \deltap{}$=163.2\,$GeV$/c$, $\sigma_{NN}=68.3$\,mb and \sigmazdc{}$=39.1\,$mb, which are loosely motivated by trying to describe the PbPb ZDC signal \cite{ZDCtoappear,ALICE:2020bta}\@.
Other parameters are based on a global analysis of PbPb collisions similar to \cite{Giacalone:2023cet}, but with values $w=0.58\,$fm, $d_\text{min}=1.00\,$fm for the nucleon width and minimum nucleon-nucleon distance (other parameters do not affect the current analysis)\@.

\section{Results}\label{sec:results}

Our main result is presented in Fig.~\ref{fig:oxygentransmutation}, which displays the cross-section of all isotopes with lifetime bigger than $1\,$s produced in \oooo{} and \nene{} collisions as a function of their momentum. Here we took $\delta p$ to be the difference in momentum with respect to the momentum needed for that isotope to circulate in the LHC beam. There is a clear peak of protons ($^{1}$H, mostly spectator nucleons) with a relatively small momentum spread that is caused by their Fermi momentum. These protons should have had half their momentum to keep circulating since their charge over mass ratio is double that of \oo{}. The peak is hence is located around -50\%.

Most of the isotopes will not circulate as currently the optics are foreseen to only allow a deviation of 0.034\% of the canonical momentum \cite{LPCpresentation}\@.
Integrating this leads to the circulating isotope production cross-sections, which are dominated by $^{4}$He. Since the cross-sections are fairly flat around $\delta p=0$ a change in optics will simply rescale the cross-sections circulating in the beam. 

It is an interesting observation that in \nene{} collisions there is relatively speaking more $^{4}$He and other isotopes than a naive cross section scaling would give. Looking carefully this is true for all small isotopes, while larger isotopes have different ratios. So in our model it is unlikely that the `bowling pin' structure of \nee{} \cite{Giacalone:2024luz} contributes significantly to the $^{4}$He production.

It is an interesting question whether the circulating isotopes can in subsequent revolutions collide with the remaining oxygen. In the beginning there will be 100\% oxygen, but slowly predominantly helium and deuterons are added to the beam, while the oxygen gets depleted due to the collisions. After about 12 hours only about half of the oxygen will be left \cite{Bruce:2021hjk}, depending a bit on optics and injector intensities. In our model, at this stage to a reasonable approximation the amount of deuteron and helium is given by the ratio of the (hadronic) cross sections, which is about $(2.108+2.219)/1420=0.3\%$.\footnote{This ignores electromagnetic depletion and other beam loss sources of the oxygen beams which reduce the transmuted fraction of the collisions proportionally.} Deuteron and helium have lower cross-sections ($0.810\,$b for \oo{}$^4$He and $0.696\,$b for \oo{}$d$), but they can have an effect on the observables at the lower range of multiplicity.

\begin{figure}[t]
    \centering
    \includegraphics[width=8cm]{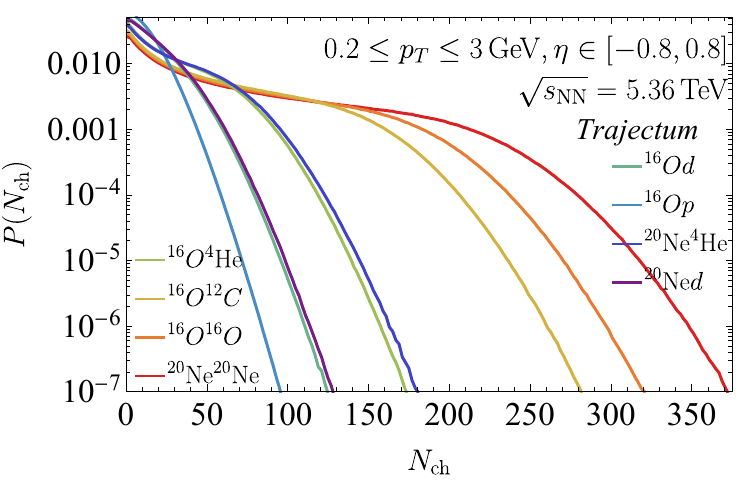}\\
    \caption{We show the $N_\text{ch}$ distributions for several light ion configurations.}
    \label{fig:ntrackdistributions}
\end{figure}

\begin{figure}[t]
    \centering
    \includegraphics[width=8cm]{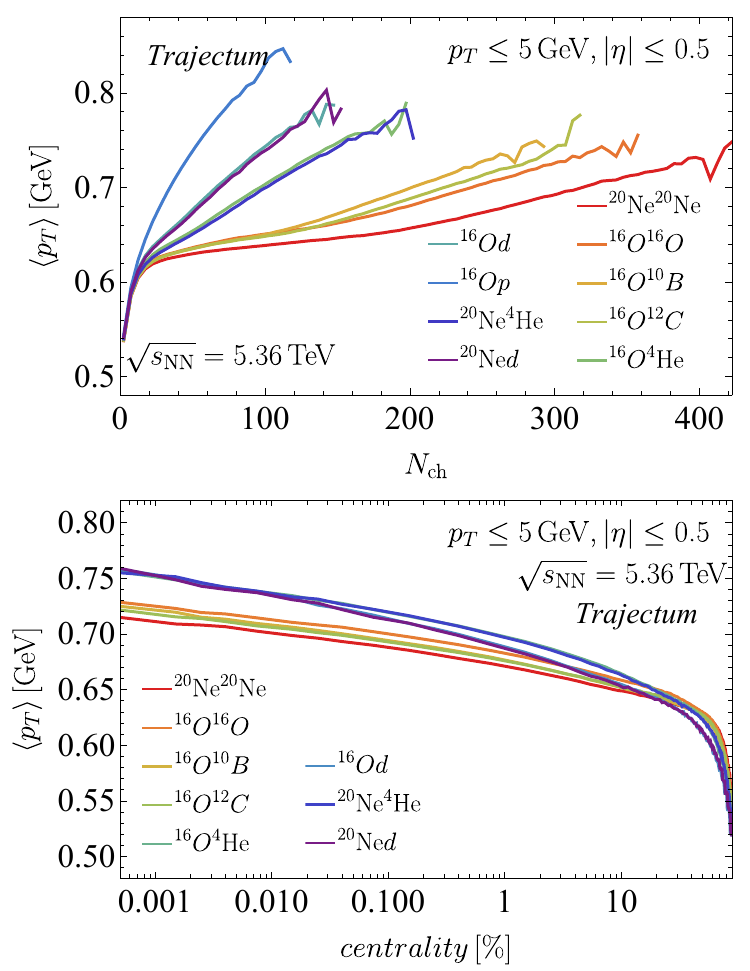}\\
    \includegraphics[width=8cm]{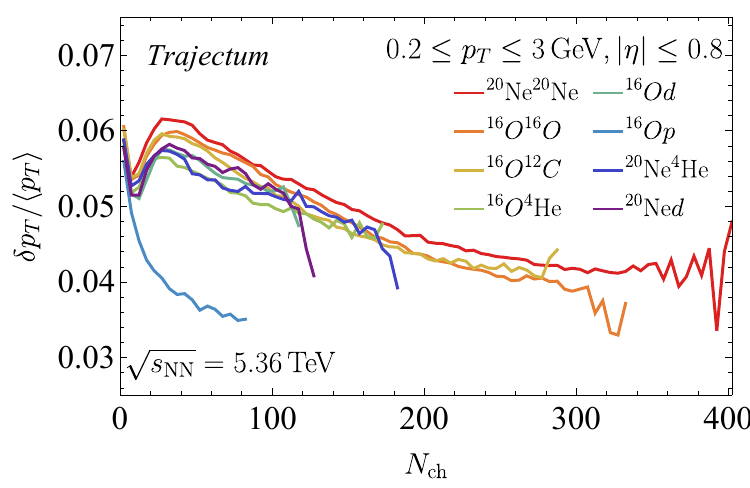}\\
\caption{We show the mean transverse momentum (top) and fluctuations thereof (bottom) as a function of $N_\text{ch}$ (top, defined as in \ref{fig:ntrackdistributions}) and centrality (bottom) for several light ion configurations. Notably there is a large difference in \mpt{} when comparing different systems at fixed multiplicity.}
    \label{fig:meanpt}
\end{figure}

\section{Discussion}

\begin{figure}[t]
    \centering
    \includegraphics[width=6cm]{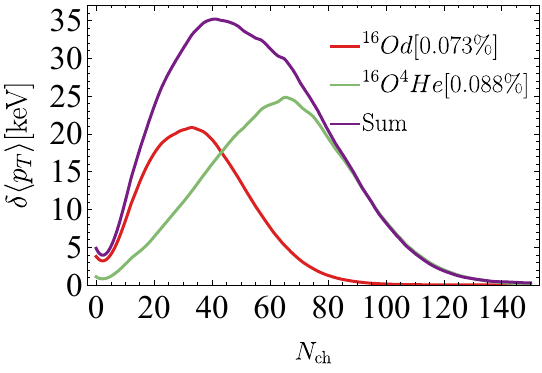}\\
    \caption{We show increase in mean $p_T$ as a function of $N_{\rm ch}$ (same cuts as in Fig.~\ref{fig:ntrackdistributions}) that would occur if 0.073\% (0.088\%) of transmuted deuteron ($^{4}$He) would be present in the beam.}
    \label{fig:deltameanptvsnch}
\end{figure}

An important uncertainty is the nucleon momenta and with it the final momenta of the clusters. In the Abrasion-Ablation Monte Carlo for Colliders (AAMMC) model the resulting momenta are for instance significantly smaller \cite{Svetlichnyi:2023nim, LPCpresentation}\@. Part of the reason is presumably that there the binding energies are the determining factor for the nucleon momenta instead of the free Fermi momenta used here. To assess this uncertainty we repeated the analysis with settings $p_\text{fermi} \rightarrow p_\text{fermi} / 2$ and \deltap{}$\rightarrow$\deltap$/2$. Results are presented in Fig.~\ref{fig:oxygentransmutationhalf} in the Supplemental Material and it can be seen that this approximately doubles the amount of most circulating isotopes. We treat this as an estimate for the systematic uncertainty.

All computations in the main text have been done using the NLEFT pinhole configurations of \oo{} isotopes. We verified, however, that no significant changes were seen using the PGCM profiles. It would also be interesting to include electromagnetic dissociation processes (dominant in e.g.~gold production from the collision of Pb ions \cite{ALICE:2024vpj})\@. For our purposes these are likely less relevant, both since the electromagnetic field of oxygen ions is much weaker and secondly since those processes carry so much momentum that the fragments likely do not stay in the beam.

The transmutation of oxygen into different elements is not only a problem as a contamination of the beam, but it presents also an opportunity. At energies available at the LHC there have only been PbPb, XeXe and $p$Pb and $pp$ collisions so far. The transmutation hence provides an opportunity to study also \oo{}$^4$He collisions at the same time as $p$O, OO and NeNe collisions are performed, much like the SMOG system at LHCb can study simultaneously fixed target collisions on beam gas. 

For this we show for several collisions systems the charged multiplicity, mean transverse momentum and transverse momentum fluctuations in Fig.~\ref{fig:ntrackdistributions}, Fig.~\ref{fig:meanpt} (top) and Fig.~\ref{fig:meanpt} (bottom) respectively (see SM for all the parameter settings). It is natural that \oo{}$^4$He collisions have a lower multiplicity, but we see that the smaller size also leads to a characteristically higher mean transverse momentum at the same multiplicity. For the case of $p$O collisions we moreover see a large difference in the transverse momentum fluctuations. The latter can be explained since the droplet size has a much more consistent size due to the round shape of the proton as compared to the other ions (especially deuteron, see SM Fig.~\ref{fig:initialrms})\@. 

It is interesting to note the similarity between \oo{}$^4$He and \nee{}$^4$He collisions, showing that both the multiplicity and the mean transverse momentum is completely determined by the shape of $^4$He. This is partly due to the self-normalisation in Fig.~\ref{fig:ntrackdistributions}, since the cross sections are different (809 and $957\,$mb respectively).
We stress that even a simple observable like the mean transverse momentum can lead to profound insights into the hydrodynamic nature of the quantum matter created in such collisions \cite{Gardim:2019brr,Nijs:2023bzv,CMS:2024sgx,ATLAS:2024jvf,ALICE:2025rtg}\@.

One of the tell-tale signs of transmutation would be the increase in the mean transverse momentum for low multiplicity collisions. In Fig.~\ref{fig:deltameanptvsnch} we show this difference for the scenario as in Section \ref{sec:results} where half of the beam is depleted by hadronic collisions. Due to the small rate of circulating nuclei the difference in $\langle p_T \rangle$ is small, but it may still be experimentally accessible since much of the systematic uncertainty cancels. With a collected sample of over 10B collisions the statistical uncertainty should also be small.

In combination with the asymmetric nature of the collision (visible in the multiplicity versus rapidity + ZDC) the difference between the original OO collisions and the transmuted collisions is likely so large that with relatively high confidence transmuted collisions can be distinguished on an event-by-event basis. On a statistical basis, using the time dependence of the multiplicity and mean transverse momentum distributions, an even more detailed understanding can likely be obtained. Differences between collision schemes of different bunches (including the foreseen four bunches that only collide in fixed target mode at SMOG) can also be exploited to track the amount of contamination. 

Significant uncertainty exists around the fragmentation region of relativistic ion collisions and the transmutation of oxygen in particular. The upcoming LHC special run hence adds one more exciting topic that we will soon learn much more about.

\section*{Acknowledgements}

We thank Roderik Bruce, Giuliano Giacalone, Jiangyong Jia, Dean Lee, Lu-Meng Liu, Raimond Snellings, Anthony Timmins, Urs Wiedemann and especially John Jowett for interesting discussions.

\emph{Trajectum} is publicly available and can be downloaded at \url{https://sites.google.com/view/govertnijs/trajectum}\@. All data files used can be found at \url{http://wilkevanderschee.nl/trajectum}. All datapoints in this paper are included also as source files.

\section*{Supplemental material}

\begin{figure*}[t]
    \centering
    \includegraphics[width=17cm]{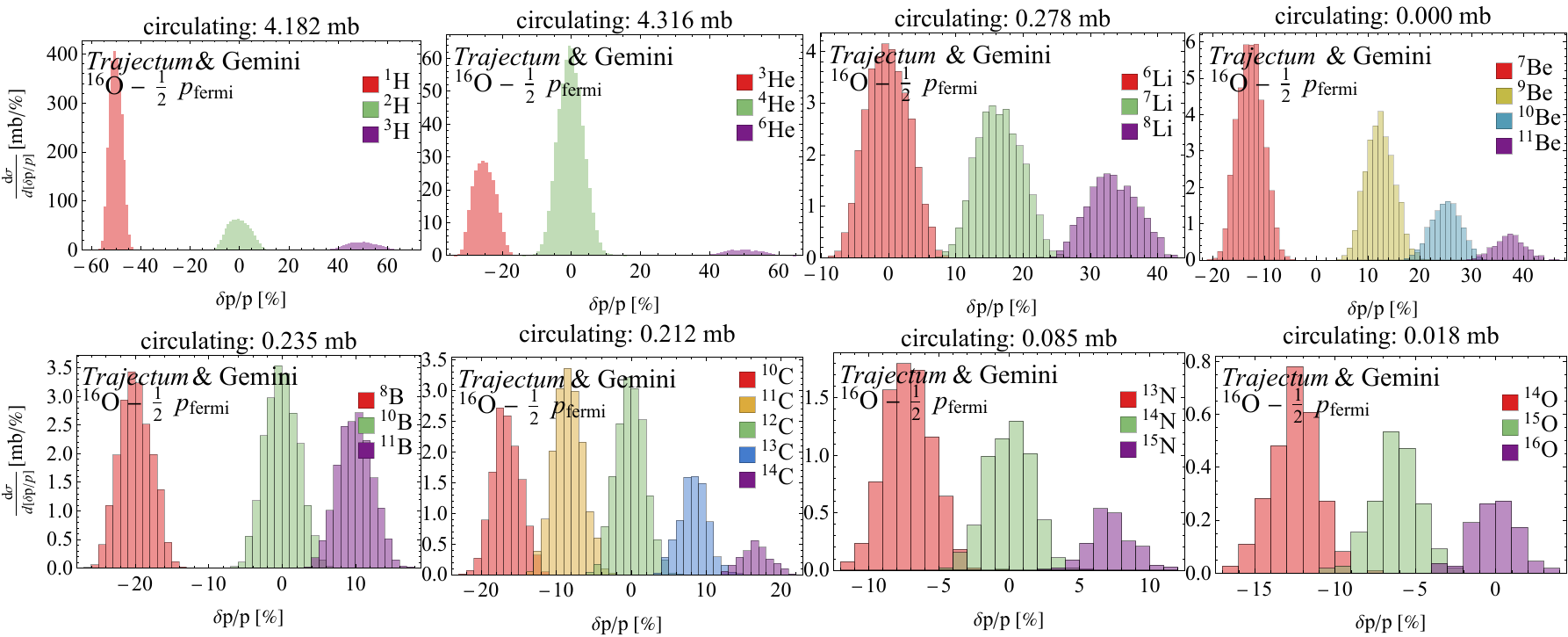}\\
    \caption{We show the equivalent figure as in \ref{fig:oxygentransmutation}, but with all initial nucleon momenta in the nucleus halved, as well as their maximum distance, \deltap{}, to be combined into a cluster. The smaller momentum leads to an approximate doubling of the transmuted isotopes in the beam.
    \label{fig:oxygentransmutationhalf}}
\end{figure*}

To study the dependence of our results on the nucleon momenta, which are uncertain, in Fig.~\ref{fig:oxygentransmutationhalf}, we repeat the computation shown in Fig.~\ref{fig:oxygentransmutation}, but with the initial nucleon momenta in each nucleus ($p_\text{fermi}$) halved, as well as the maximum distance to combine into a cluster (\deltap{})\@.
Doing this narrows the momentum distribution of each species, and as such the circulating isotope production cross-section is larger by about a factor 2 for the most prevalently produced isotopes.

\begin{figure}[t]
    \centering
    \includegraphics[width=8cm]{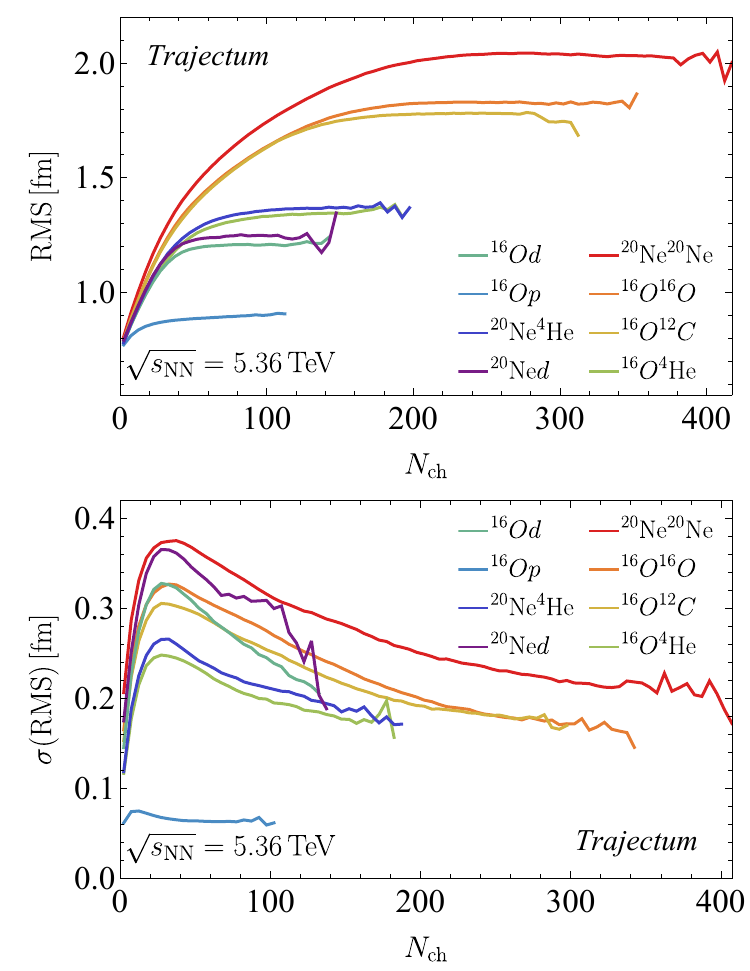}\\
    \caption{We show the initial size (top) and its standard deviation (bottom) of the QGP created for the various systems studied at the initial starting time of hydrodynamics at $\tau = 0.54\,$fm$/c$\@.
    \label{fig:initialrms}}
\end{figure}

Even though this is not directly observable, in Fig.~\ref{fig:initialrms} we show the initial root mean squared size (RMS) of the QGP when it is produced (top), as well as the event-by-event standard deviation of this RMS (bottom)\@. As expected, the produced size of the QGP scales with the size of the projectiles. It is also interesting to note that the $\sigma(\text{RMS})$ of $d^{16}$O is larger than that of $^4$He$^{16}$O\@. The reason for this is that the elongated shape of the deuteron causes large fluctuations in the size of the QGP for different orientations of the deuteron.

\begin{figure*}[t]
    \centering
    \includegraphics[width=17cm]{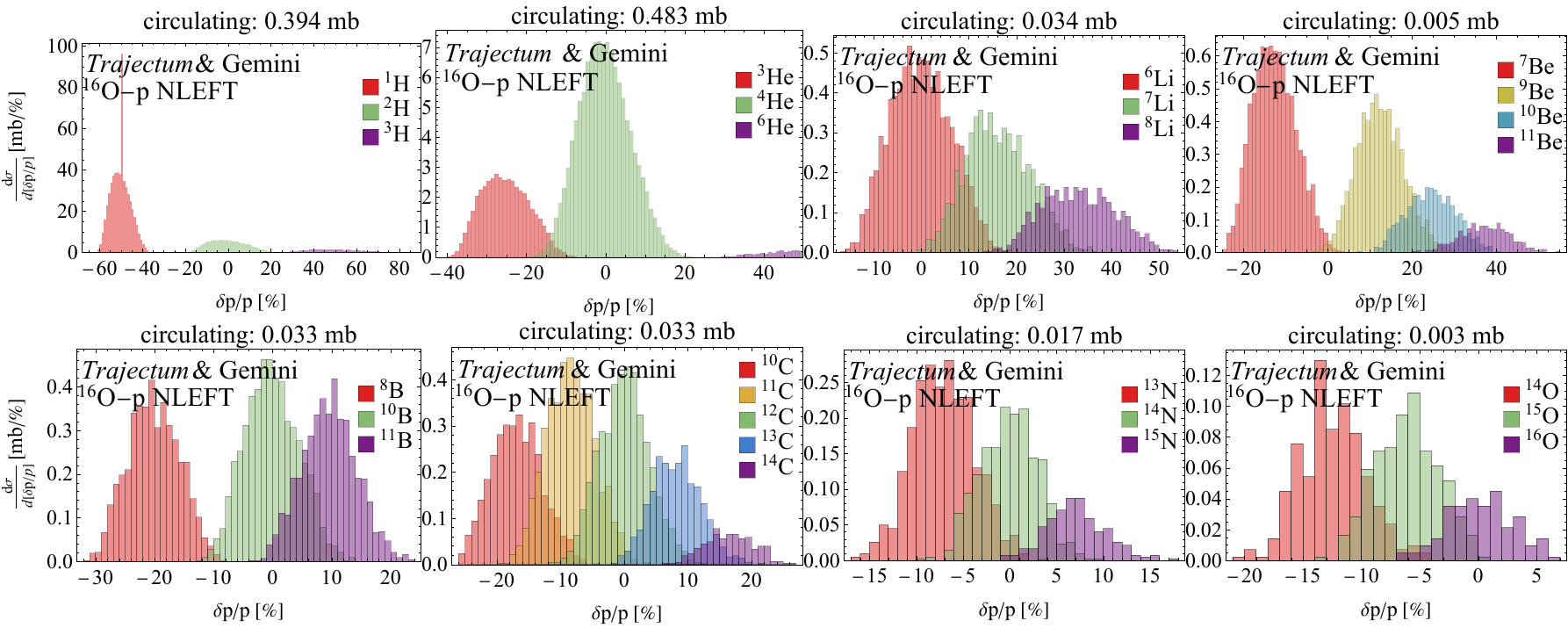}\\
    \caption{We show the equivalent figure to Fig.~\ref{fig:oxygentransmutation}, but for $p$\oo{} collisions. The total hadronic $p$\oo{} cross-section is about $0.472\,$b for our model.
    \label{fig:poxygentransmutation}}
\end{figure*}

\begin{table}[t]
\centering
 \begin{tabular}{ c c  } 
parameter & value \\
 \hline
Norm [5.36 TeV] & 22.9 \\
$w$ (fm) & 0.578 \\
cent$_{\rm norm}$ (\%) & 97.9 \\
$\sigma_{\rm fluct}$ & 0.511 \\
$p$ & -0.079 \\
$q$ & 1.25 \\
$d_{\rm min}$ (fm) & 0.997 \\
$T_{\rm switch}$ (MeV) & 149.7 \\
$n_{\rm c}$ & 2.30 \\
$v/w$ & 0.864 \\
$\tau_{\rm hyd}$ (fm/$c$) & 0.544 \\
$(\eta/s)_{\rm 0.15~GeV}$ & 0.170 \\
$(\eta/s)_{\rm 0.3~GeV}$ (GeV$^{-1}$) & 0.190 \\
$(\eta/s)_{\delta\rm 0.5~GeV}$ (GeV$^{-1}$) & 0.291 \\
$(\eta/s)_{\rm 0.8~GeV}$ & 0.324 \\
$(\zeta/s)_{\rm max}$ & 0.0372 \\
$(\zeta/s)_{\rm width}$ & 0.138 \\
$(\zeta/s)_{T_0}$ & 0.356 \\
$\tau_\pi T$ & 1.66 \\
$\tau_{\pi\pi}/\tau_\pi$ & 3.48 \\
$r_{\rm hyd}$ & 0.829 \\
$a_{\rm EOS}$ & -9.41 \\ 
$f_{\rm SMASH}$ & 0.903 \\
\hline
\end{tabular}
\caption{List of all model parameters constituting the hydrodynamic model used in this paper. Only the nucleon width $w$ and the minimum distance $d_{\rm min}$ affect the isotope production presented as the main result, but the other parameters influence the basic observables as also presented.}
\label{tab:1}
\end{table}

\bibliographystyle{apsrev4-1}
\bibliography{transmutation, manual}

\begin{thebibliography}{24}%
\makeatletter
\providecommand \@ifxundefined [1]{%
 \@ifx{#1\undefined}
}%
\providecommand \@ifnum [1]{%
 \ifnum #1\expandafter \@firstoftwo
 \else \expandafter \@secondoftwo
 \fi
}%
\providecommand \@ifx [1]{%
 \ifx #1\expandafter \@firstoftwo
 \else \expandafter \@secondoftwo
 \fi
}%
\providecommand \natexlab [1]{#1}%
\providecommand \enquote  [1]{``#1''}%
\providecommand \bibnamefont  [1]{#1}%
\providecommand \bibfnamefont [1]{#1}%
\providecommand \citenamefont [1]{#1}%
\providecommand \href@noop [0]{\@secondoftwo}%
\providecommand \href [0]{\begingroup \@sanitize@url \@href}%
\providecommand \@href[1]{\@@startlink{#1}\@@href}%
\providecommand \@@href[1]{\endgroup#1\@@endlink}%
\providecommand \@sanitize@url [0]{\catcode `\\12\catcode `\$12\catcode `\&12\catcode `\#12\catcode `\^12\catcode `\_12\catcode `\%12\relax}%
\providecommand \@@startlink[1]{}%
\providecommand \@@endlink[0]{}%
\providecommand \url  [0]{\begingroup\@sanitize@url \@url }%
\providecommand \@url [1]{\endgroup\@href {#1}{\urlprefix }}%
\providecommand \urlprefix  [0]{URL }%
\providecommand \Eprint [0]{\href }%
\providecommand \doibase [0]{http://dx.doi.org/}%
\providecommand \selectlanguage [0]{\@gobble}%
\providecommand \bibinfo  [0]{\@secondoftwo}%
\providecommand \bibfield  [0]{\@secondoftwo}%
\providecommand \translation [1]{[#1]}%
\providecommand \BibitemOpen [0]{}%
\providecommand \bibitemStop [0]{}%
\providecommand \bibitemNoStop [0]{.\EOS\space}%
\providecommand \EOS [0]{\spacefactor3000\relax}%
\providecommand \BibitemShut  [1]{\csname bibitem#1\endcsname}%
\let\auto@bib@innerbib\@empty
\bibitem [{\citenamefont {Brewer}\ \emph {et~al.}(2021)\citenamefont {Brewer}, \citenamefont {Mazeliauskas},\ and\ \citenamefont {van~der Schee}}]{Brewer:2021kiv}%
  \BibitemOpen
  \bibfield  {author} {\bibinfo {author} {\bibfnamefont {J.}~\bibnamefont {Brewer}}, \bibinfo {author} {\bibfnamefont {A.}~\bibnamefont {Mazeliauskas}}, \ and\ \bibinfo {author} {\bibfnamefont {W.}~\bibnamefont {van~der Schee}},\ }in\ \href@noop {} {\emph {\bibinfo {booktitle} {{Opportunities of OO and pO collisions at the LHC}}}}\ (\bibinfo {year} {2021})\ \Eprint {http://arxiv.org/abs/2103.01939} {arXiv:2103.01939 [hep-ph]} \BibitemShut {NoStop}%
\bibitem [{\citenamefont {Grosse-Oetringhaus}\ and\ \citenamefont {Wiedemann}(2024)}]{Grosse-Oetringhaus:2024bwr}%
  \BibitemOpen
  \bibfield  {author} {\bibinfo {author} {\bibfnamefont {J.~F.}\ \bibnamefont {Grosse-Oetringhaus}}\ and\ \bibinfo {author} {\bibfnamefont {U.~A.}\ \bibnamefont {Wiedemann}},\ }\href@noop {} {\  (\bibinfo {year} {2024})},\ \Eprint {http://arxiv.org/abs/2407.07484} {arXiv:2407.07484 [hep-ex]} \BibitemShut {NoStop}%
\bibitem [{\citenamefont {Giacalone}\ \emph {et~al.}(2024)\citenamefont {Giacalone} \emph {et~al.}}]{Giacalone:2024luz}%
  \BibitemOpen
  \bibfield  {author} {\bibinfo {author} {\bibfnamefont {G.}~\bibnamefont {Giacalone}} \emph {et~al.},\ }\href@noop {} {\  (\bibinfo {year} {2024})},\ \Eprint {http://arxiv.org/abs/2402.05995} {arXiv:2402.05995 [nucl-th]} \BibitemShut {NoStop}%
\bibitem [{\citenamefont {Giacalone}\ \emph {et~al.}(2025)\citenamefont {Giacalone} \emph {et~al.}}]{Giacalone:2024ixe}%
  \BibitemOpen
  \bibfield  {author} {\bibinfo {author} {\bibfnamefont {G.}~\bibnamefont {Giacalone}} \emph {et~al.},\ }\href {\doibase 10.1103/PhysRevLett.134.082301} {\bibfield  {journal} {\bibinfo  {journal} {Phys. Rev. Lett.}\ }\textbf {\bibinfo {volume} {134}},\ \bibinfo {pages} {082301} (\bibinfo {year} {2025})},\ \Eprint {http://arxiv.org/abs/2405.20210} {arXiv:2405.20210 [nucl-th]} \BibitemShut {NoStop}%
\bibitem [{\citenamefont {Summerfield}\ \emph {et~al.}(2021)\citenamefont {Summerfield}, \citenamefont {Lu}, \citenamefont {Plumberg}, \citenamefont {Lee}, \citenamefont {Noronha-Hostler},\ and\ \citenamefont {Timmins}}]{Summerfield:2021oex}%
  \BibitemOpen
  \bibfield  {author} {\bibinfo {author} {\bibfnamefont {N.}~\bibnamefont {Summerfield}}, \bibinfo {author} {\bibfnamefont {B.-N.}\ \bibnamefont {Lu}}, \bibinfo {author} {\bibfnamefont {C.}~\bibnamefont {Plumberg}}, \bibinfo {author} {\bibfnamefont {D.}~\bibnamefont {Lee}}, \bibinfo {author} {\bibfnamefont {J.}~\bibnamefont {Noronha-Hostler}}, \ and\ \bibinfo {author} {\bibfnamefont {A.}~\bibnamefont {Timmins}},\ }\href {\doibase 10.1103/PhysRevC.104.L041901} {\bibfield  {journal} {\bibinfo  {journal} {Phys. Rev. C}\ }\textbf {\bibinfo {volume} {104}},\ \bibinfo {pages} {L041901} (\bibinfo {year} {2021})},\ \Eprint {http://arxiv.org/abs/2103.03345} {arXiv:2103.03345 [nucl-th]} \BibitemShut {NoStop}%
\bibitem [{\citenamefont {Wang}\ \emph {et~al.}(2024)\citenamefont {Wang}, \citenamefont {Zhao}, \citenamefont {Cao}, \citenamefont {Xu},\ and\ \citenamefont {Song}}]{YuanyuanWang:2024sgp}%
  \BibitemOpen
  \bibfield  {author} {\bibinfo {author} {\bibfnamefont {Y.}~\bibnamefont {Wang}}, \bibinfo {author} {\bibfnamefont {S.}~\bibnamefont {Zhao}}, \bibinfo {author} {\bibfnamefont {B.}~\bibnamefont {Cao}}, \bibinfo {author} {\bibfnamefont {H.-j.}\ \bibnamefont {Xu}}, \ and\ \bibinfo {author} {\bibfnamefont {H.}~\bibnamefont {Song}},\ }\href {\doibase 10.1103/PhysRevC.109.L051904} {\bibfield  {journal} {\bibinfo  {journal} {Phys. Rev. C}\ }\textbf {\bibinfo {volume} {109}},\ \bibinfo {pages} {L051904} (\bibinfo {year} {2024})},\ \Eprint {http://arxiv.org/abs/2401.15723} {arXiv:2401.15723 [nucl-th]} \BibitemShut {NoStop}%
\bibitem [{\citenamefont {Liu}\ \emph {et~al.}(2022)\citenamefont {Liu}, \citenamefont {Zhang}, \citenamefont {Xu}, \citenamefont {Jia},\ and\ \citenamefont {Peng}}]{Liu:2022xlm}%
  \BibitemOpen
  \bibfield  {author} {\bibinfo {author} {\bibfnamefont {L.-M.}\ \bibnamefont {Liu}}, \bibinfo {author} {\bibfnamefont {C.-J.}\ \bibnamefont {Zhang}}, \bibinfo {author} {\bibfnamefont {J.}~\bibnamefont {Xu}}, \bibinfo {author} {\bibfnamefont {J.}~\bibnamefont {Jia}}, \ and\ \bibinfo {author} {\bibfnamefont {G.-X.}\ \bibnamefont {Peng}},\ }\href {\doibase 10.1103/PhysRevC.106.034913} {\bibfield  {journal} {\bibinfo  {journal} {Phys. Rev. C}\ }\textbf {\bibinfo {volume} {106}},\ \bibinfo {pages} {034913} (\bibinfo {year} {2022})},\ \Eprint {http://arxiv.org/abs/2209.03106} {arXiv:2209.03106 [nucl-th]} \BibitemShut {NoStop}%
\bibitem [{\citenamefont {Liu}\ \emph {et~al.}(2024)\citenamefont {Liu}, \citenamefont {Li}, \citenamefont {Wang}, \citenamefont {Xu}, \citenamefont {Ren},\ and\ \citenamefont {Huang}}]{Liu:2023gun}%
  \BibitemOpen
  \bibfield  {author} {\bibinfo {author} {\bibfnamefont {L.-M.}\ \bibnamefont {Liu}}, \bibinfo {author} {\bibfnamefont {S.-J.}\ \bibnamefont {Li}}, \bibinfo {author} {\bibfnamefont {Z.}~\bibnamefont {Wang}}, \bibinfo {author} {\bibfnamefont {J.}~\bibnamefont {Xu}}, \bibinfo {author} {\bibfnamefont {Z.-Z.}\ \bibnamefont {Ren}}, \ and\ \bibinfo {author} {\bibfnamefont {X.-G.}\ \bibnamefont {Huang}},\ }\href {\doibase 10.1016/j.physletb.2024.138724} {\bibfield  {journal} {\bibinfo  {journal} {Phys. Lett. B}\ }\textbf {\bibinfo {volume} {854}},\ \bibinfo {pages} {138724} (\bibinfo {year} {2024})},\ \Eprint {http://arxiv.org/abs/2312.13572} {arXiv:2312.13572 [nucl-th]} \BibitemShut {NoStop}%
\bibitem [{\citenamefont {Liu}\ \emph {et~al.}(2023)\citenamefont {Liu}, \citenamefont {Xu},\ and\ \citenamefont {Peng}}]{Liu:2023qeq}%
  \BibitemOpen
  \bibfield  {author} {\bibinfo {author} {\bibfnamefont {L.-M.}\ \bibnamefont {Liu}}, \bibinfo {author} {\bibfnamefont {J.}~\bibnamefont {Xu}}, \ and\ \bibinfo {author} {\bibfnamefont {G.-X.}\ \bibnamefont {Peng}},\ }\href {\doibase 10.1016/j.physletb.2023.137701} {\bibfield  {journal} {\bibinfo  {journal} {Phys. Lett. B}\ }\textbf {\bibinfo {volume} {838}},\ \bibinfo {pages} {137701} (\bibinfo {year} {2023})},\ \Eprint {http://arxiv.org/abs/2301.07893} {arXiv:2301.07893 [nucl-th]} \BibitemShut {NoStop}%
\bibitem [{\citenamefont {Nijs}\ and\ \citenamefont {van~der Schee}(2023)}]{Nijs:2023yab}%
  \BibitemOpen
  \bibfield  {author} {\bibinfo {author} {\bibfnamefont {G.}~\bibnamefont {Nijs}}\ and\ \bibinfo {author} {\bibfnamefont {W.}~\bibnamefont {van~der Schee}},\ }\href@noop {} {\  (\bibinfo {year} {2023})},\ \Eprint {http://arxiv.org/abs/2304.06191} {arXiv:2304.06191 [nucl-th]} \BibitemShut {NoStop}%
\bibitem [{\citenamefont {Nijs}\ and\ \citenamefont {van~der Schee}()}]{ZDCtoappear}%
  \BibitemOpen
  \bibfield  {author} {\bibinfo {author} {\bibfnamefont {G.}~\bibnamefont {Nijs}}\ and\ \bibinfo {author} {\bibfnamefont {W.}~\bibnamefont {van~der Schee}},\ }\href@noop {} {}\bibinfo {note} {To appear}\BibitemShut {NoStop}%
\bibitem [{\citenamefont {Charity}\ \emph {et~al.}(1988)\citenamefont {Charity} \emph {et~al.}}]{Charity:1988zz}%
  \BibitemOpen
  \bibfield  {author} {\bibinfo {author} {\bibfnamefont {R.~J.}\ \bibnamefont {Charity}} \emph {et~al.},\ }\href {\doibase 10.1016/0375-9474(88)90542-8} {\bibfield  {journal} {\bibinfo  {journal} {Nucl. Phys. A}\ }\textbf {\bibinfo {volume} {483}},\ \bibinfo {pages} {371} (\bibinfo {year} {1988})}\BibitemShut {NoStop}%
\bibitem [{\citenamefont {Charity}(2010)}]{Charity:2010wk}%
  \BibitemOpen
  \bibfield  {author} {\bibinfo {author} {\bibfnamefont {R.~J.}\ \bibnamefont {Charity}},\ }\href {\doibase 10.1103/PhysRevC.82.014610} {\bibfield  {journal} {\bibinfo  {journal} {Phys. Rev. C}\ }\textbf {\bibinfo {volume} {82}},\ \bibinfo {pages} {014610} (\bibinfo {year} {2010})},\ \Eprint {http://arxiv.org/abs/1006.5018} {arXiv:1006.5018 [nucl-th]} \BibitemShut {NoStop}%
\bibitem [{\citenamefont {ALICE}(2020)}]{ALICE:2020bta}%
  \BibitemOpen
  \bibfield  {author} {\bibinfo {author} {\bibnamefont {ALICE}},\ }\href@noop {} {\enquote {\bibinfo {title} {{Data-driven model for the emission of spectator nucleons as a function of centrality in Pb-Pb collisions at LHC energies}},}\ }\bibinfo {howpublished} {ALICE-PUBLIC-2020-001} (\bibinfo {year} {2020})\BibitemShut {NoStop}%
\bibitem [{\citenamefont {Giacalone}\ \emph {et~al.}(2023)\citenamefont {Giacalone}, \citenamefont {Nijs},\ and\ \citenamefont {van~der Schee}}]{Giacalone:2023cet}%
  \BibitemOpen
  \bibfield  {author} {\bibinfo {author} {\bibfnamefont {G.}~\bibnamefont {Giacalone}}, \bibinfo {author} {\bibfnamefont {G.}~\bibnamefont {Nijs}}, \ and\ \bibinfo {author} {\bibfnamefont {W.}~\bibnamefont {van~der Schee}},\ }\href {\doibase 10.1103/PhysRevLett.131.202302} {\bibfield  {journal} {\bibinfo  {journal} {Phys. Rev. Lett.}\ }\textbf {\bibinfo {volume} {131}},\ \bibinfo {pages} {202302} (\bibinfo {year} {2023})},\ \Eprint {http://arxiv.org/abs/2305.00015} {arXiv:2305.00015 [nucl-th]} \BibitemShut {NoStop}%
\bibitem [{\citenamefont {Jowett}(2025)}]{LPCpresentation}%
  \BibitemOpen
  \bibfield  {author} {\bibinfo {author} {\bibfnamefont {J.}~\bibnamefont {Jowett}},\ }\href@noop {} {} (\bibinfo {year} {2025}),\ \bibinfo {note} {{Presentation at LPC 26/6/2025}}\BibitemShut {NoStop}%
\bibitem [{\citenamefont {Bruce}\ \emph {et~al.}(2021)\citenamefont {Bruce}, \citenamefont {Alemany-Fern\'andez}, \citenamefont {Bartosik}, \citenamefont {Jebramcik}, \citenamefont {Jowett},\ and\ \citenamefont {Schaumann}}]{Bruce:2021hjk}%
  \BibitemOpen
  \bibfield  {author} {\bibinfo {author} {\bibfnamefont {R.}~\bibnamefont {Bruce}}, \bibinfo {author} {\bibfnamefont {R.}~\bibnamefont {Alemany-Fern\'andez}}, \bibinfo {author} {\bibfnamefont {H.}~\bibnamefont {Bartosik}}, \bibinfo {author} {\bibfnamefont {M.}~\bibnamefont {Jebramcik}}, \bibinfo {author} {\bibfnamefont {J.}~\bibnamefont {Jowett}}, \ and\ \bibinfo {author} {\bibfnamefont {M.}~\bibnamefont {Schaumann}},\ }in\ \href {\doibase 10.18429/JACoW-IPAC2021-MOPAB005} {\emph {\bibinfo {booktitle} {{12th International Particle Accelerator Conference~}}}}\ (\bibinfo {year} {2021})\BibitemShut {NoStop}%
\bibitem [{\citenamefont {Svetlichnyi}\ \emph {et~al.}(2023)\citenamefont {Svetlichnyi}, \citenamefont {Savenkov}, \citenamefont {Nepeivoda},\ and\ \citenamefont {Pshenichnov}}]{Svetlichnyi:2023nim}%
  \BibitemOpen
  \bibfield  {author} {\bibinfo {author} {\bibfnamefont {A.}~\bibnamefont {Svetlichnyi}}, \bibinfo {author} {\bibfnamefont {S.}~\bibnamefont {Savenkov}}, \bibinfo {author} {\bibfnamefont {R.}~\bibnamefont {Nepeivoda}}, \ and\ \bibinfo {author} {\bibfnamefont {I.}~\bibnamefont {Pshenichnov}},\ }\href {\doibase 10.3390/physics5020027} {\bibfield  {journal} {\bibinfo  {journal} {MDPI Physics}\ }\textbf {\bibinfo {volume} {5}},\ \bibinfo {pages} {381} (\bibinfo {year} {2023})}\BibitemShut {NoStop}%
\bibitem [{\citenamefont {Acharya}\ \emph {et~al.}(2025)\citenamefont {Acharya} \emph {et~al.}}]{ALICE:2024vpj}%
  \BibitemOpen
  \bibfield  {author} {\bibinfo {author} {\bibfnamefont {S.}~\bibnamefont {Acharya}} \emph {et~al.} (\bibinfo {collaboration} {ALICE}),\ }\href {\doibase 10.1103/PhysRevC.111.054906} {\bibfield  {journal} {\bibinfo  {journal} {Phys. Rev. C}\ }\textbf {\bibinfo {volume} {111}},\ \bibinfo {pages} {054906} (\bibinfo {year} {2025})},\ \Eprint {http://arxiv.org/abs/2411.07058} {arXiv:2411.07058 [nucl-ex]} \BibitemShut {NoStop}%
\bibitem [{\citenamefont {Gardim}\ \emph {et~al.}(2020)\citenamefont {Gardim}, \citenamefont {Giacalone},\ and\ \citenamefont {Ollitrault}}]{Gardim:2019brr}%
  \BibitemOpen
  \bibfield  {author} {\bibinfo {author} {\bibfnamefont {F.~G.}\ \bibnamefont {Gardim}}, \bibinfo {author} {\bibfnamefont {G.}~\bibnamefont {Giacalone}}, \ and\ \bibinfo {author} {\bibfnamefont {J.-Y.}\ \bibnamefont {Ollitrault}},\ }\href {\doibase 10.1016/j.physletb.2020.135749} {\bibfield  {journal} {\bibinfo  {journal} {Phys. Lett. B}\ }\textbf {\bibinfo {volume} {809}},\ \bibinfo {pages} {135749} (\bibinfo {year} {2020})},\ \Eprint {http://arxiv.org/abs/1909.11609} {arXiv:1909.11609 [nucl-th]} \BibitemShut {NoStop}%
\bibitem [{\citenamefont {Nijs}\ and\ \citenamefont {van~der Schee}(2024)}]{Nijs:2023bzv}%
  \BibitemOpen
  \bibfield  {author} {\bibinfo {author} {\bibfnamefont {G.}~\bibnamefont {Nijs}}\ and\ \bibinfo {author} {\bibfnamefont {W.}~\bibnamefont {van~der Schee}},\ }\href {\doibase 10.1016/j.physletb.2024.138636} {\bibfield  {journal} {\bibinfo  {journal} {Phys. Lett. B}\ }\textbf {\bibinfo {volume} {853}},\ \bibinfo {pages} {138636} (\bibinfo {year} {2024})},\ \Eprint {http://arxiv.org/abs/2312.04623} {arXiv:2312.04623 [nucl-th]} \BibitemShut {NoStop}%
\bibitem [{\citenamefont {Hayrapetyan}\ \emph {et~al.}(2024)\citenamefont {Hayrapetyan} \emph {et~al.}}]{CMS:2024sgx}%
  \BibitemOpen
  \bibfield  {author} {\bibinfo {author} {\bibfnamefont {A.}~\bibnamefont {Hayrapetyan}} \emph {et~al.} (\bibinfo {collaboration} {CMS}),\ }\href {\doibase 10.1088/1361-6633/ad4b9b} {\bibfield  {journal} {\bibinfo  {journal} {Rept. Prog. Phys.}\ }\textbf {\bibinfo {volume} {87}},\ \bibinfo {pages} {077801} (\bibinfo {year} {2024})},\ \Eprint {http://arxiv.org/abs/2401.06896} {arXiv:2401.06896 [nucl-ex]} \BibitemShut {NoStop}%
\bibitem [{\citenamefont {Aad}\ \emph {et~al.}(2024)\citenamefont {Aad} \emph {et~al.}}]{ATLAS:2024jvf}%
  \BibitemOpen
  \bibfield  {author} {\bibinfo {author} {\bibfnamefont {G.}~\bibnamefont {Aad}} \emph {et~al.} (\bibinfo {collaboration} {ATLAS}),\ }\href {\doibase 10.1103/PhysRevLett.133.252301} {\bibfield  {journal} {\bibinfo  {journal} {Phys. Rev. Lett.}\ }\textbf {\bibinfo {volume} {133}},\ \bibinfo {pages} {252301} (\bibinfo {year} {2024})},\ \Eprint {http://arxiv.org/abs/2407.06413} {arXiv:2407.06413 [nucl-ex]} \BibitemShut {NoStop}%
\bibitem [{\citenamefont {Abualrob}\ \emph {et~al.}(2025)\citenamefont {Abualrob} \emph {et~al.}}]{ALICE:2025rtg}%
  \BibitemOpen
  \bibfield  {author} {\bibinfo {author} {\bibfnamefont {I.~J.}\ \bibnamefont {Abualrob}} \emph {et~al.} (\bibinfo {collaboration} {ALICE}),\ }\href@noop {} {\  (\bibinfo {year} {2025})},\ \Eprint {http://arxiv.org/abs/2506.10394} {arXiv:2506.10394 [nucl-ex]} \BibitemShut {NoStop}%
\end{thebibliography}%

\end{document}